# Source-region calculation of dipole power

F. S. Felber[a]

*Physics Division, Starmark, Inc., P. O. Box 270710, San Diego, California 92198*

An exact calculation of the retarded electric field in the source region of a system of individual charges, expanded to third order in velocity, shows that all nonrelativistic accelerated charges in a system emit dipole electromagnetic energy at the Larmor rate. In general, some of the emitted dipole power is reabsorbed by doing work on other charges in the system, and some is radiated. If the system has zero net dipole moment, all the emitted dipole power is reabsorbed, and none is radiated.

## I. INTRODUCTION

A solitary accelerating charge produces dipole (and higher order multipole) electromagnetic radiation. But a system of neighboring accelerating charges that has zero net electric dipole moment produces no dipole radiation.

Each accelerating charge in a system of charges has no information about the instantaneous net electric dipole moment of the system. At every instant, therefore, each accelerated charge must expend electromagnetic energy at a rate that depends only on its own acceleration, and that is independent of the acceleration of other charges in the system. Each accelerated charge expends electromagnetic energy at the Larmor rate. This paper analyzes what happens to that electromagnetic power after it is <u>emitted</u>, that is, whether it becomes <u>radiated</u> or <u>absorbed</u>.

What happens to the dipole electromagnetic power emitted by each accelerating charge in a system with zero net dipole moment? Does radiated power become undetectable or unphysical by destructive interference? Or is the power somehow switched off at the source? Or, as has been suggested for a different context [1], does electromagnetic energy flow from large radius into the source region? This paper provides an answer based wholly on physics in the source region, an answer that avoids the usual explanations of interference of fields in the far (radiation) zone.

Equally of interest is what happens when $N$ parallel, identical neighboring dipoles undergo an identical acceleration. Then the total dipole electromagnetic power is not $N$ times the power radiated by a solitary dipole, but $N^2$. From where does the extra electromagnetic power come?

This paper calculates the work done by accelerating charges on each other in a simple system having zero net dipole moment. The work done on each charge accounts for the vanishing dipole radiation from each charge. And when the dipole moment of a simple system of charges is doubled, then the work done by the charges accounts for a quadrupling of power radiated from the system.

These insights into the question of vanishing dipole radiation may not seem surprising. Perhaps more surprising, though, is that the *nonrelativistic*, velocity-independent Larmor power vanishes from a system having zero dipole moment owing to a *high-order relativistic correction* to the electric field.

In the nonrelativistic limit, the Larmor (dipole) radiation power is known to be independent of particle velocity. The instantaneous Larmor power from an accelerating charge is about the same for a motionless charge as for one traveling at one-tenth the speed of light. Yet, in order to show that the work done by a system of charges with zero net dipole moment accounts for vanishing dipole radiation, we must calculate the rate of work to order $\beta^4$, where $\beta$ is the ratio of particle speed to speed of light, $c$.

The Larmor power radiated by a <u>solitary</u> accelerating charge can be derived simply from first principles within a factor of order unity. Consider two identical masses that undergo identical accelerations, but one is charged and the other is not. A greater force must be applied to the charged mass to produce the same acceleration, because the electrostatic field energy of the charged mass must be accelerated with the mass [2–4].

Suppose a charge $q$ is given an acceleration $a$ during a brief time $\Delta t$. The field that is accelerated during $\Delta t$ extends to a radius $c\Delta t$. The electrostatic field energy that has not yet been accelerated during $\Delta t$ is of order $\varepsilon \approx q^2/(c\Delta t)$. This unaccelerated electrostatic energy has an equivalent mass $m \approx \varepsilon/c^2$. The work to be done by the radiation field on this equivalent mass to accelerate it to speed $a\Delta t$ is about $m(a\Delta t)^2/2$. And since this radiation was produced in a time $\Delta t$, the radiation power is about $P \approx m(a\Delta t)^2/2\Delta t \approx q^2 a^2/2c^3$, which is about equal to the actual radiated Larmor power [5],

$$P_L = 2q^2 a^2/3c^3. \qquad (1)$$

Viewed in this way, Larmor radiation is seen to be an accounting mechanism that adjusts electromagnetic fields at a distance to correspond to the state of their source. This approximate derivation is discussed further in Sec. V.

## II. MODEL

Next, we consider a system of charges in which the magnitudes of both the accelerations and the velocities of the charges remain constant. The charges are caused to move in circular orbits of radius $b$ at constant angular frequency $\omega$. The constant magnitude of centripetal acceleration is $\omega^2 b$. The constant magnitude of tangential velocity is $\omega b$.

This model is useful in its simplicity. Such a model is simpler to treat than a model of parallel, coherently oscillating electric dipoles, owing to the constant magnitudes in uniform circular motion of such quantities as ranges, speeds, accelerations, kinetic energies, rates of work done on other charges, radiated powers, etc.

A similar model of symmetrically distributed charges in uniform circular motion has been used for related calculations. The model was used to calculate radiated power and the radiative reaction force to low orders in $\beta$ [6], and to calculate the exact radiated power in certain external electric and magnetic fields [7]. In this paper, the relativistic model is used to calculate the exact rate of work done by one charge on another, and to show how the dipole electromagnetic energy expended by each accelerating charge is partitioned between radiation and work done on other charges in the system.

---

[a] Electronic mail: Starmark@san.rr.com

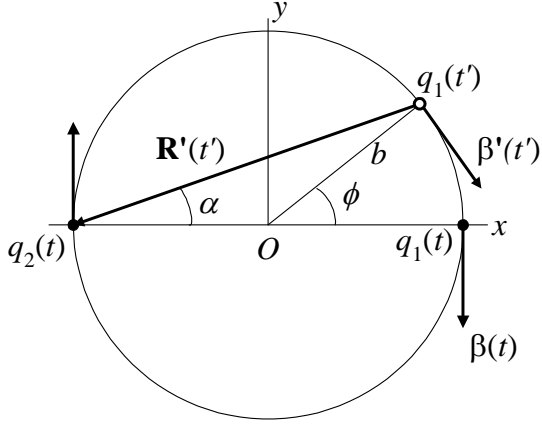

FIG. 1. Geometry of a system of charges $q_1$ and $q_2$ revolving on opposite sides of a circular orbit. Charge $q_1$ is also shown at the retarded time $t'$ with respect to $q_2(t)$.

As shown by a textbook calculation [8], and later emphasized for just this sort of model [9], care must be taken in calculating the radiated power of multipoles higher than dipole. Relativistic effects contribute to radiation at higher order terms in $\beta$ not only through harmonics in the fundamental frequency $\omega$, but also in relativistic modifications to the nonrelativistic multipole-radiation formulas [9], like Eq. (1). This caveat does not apply to the rate of work calculated in this paper of charges on other charges, which is correct to all orders of $\beta$, and does not apply to radiation calculations if one considers only lowest-order dipole radiation, which is sufficient for the purposes of this paper.

Many approaches can be used to show what ultimately happens to the electromagnetic power emitted by accelerating charges. For example, a general derivation of electromagnetic conservation laws has been based on the freedom to consider charge and current sources subdivided in different ways [10]. For the particular calculation of this paper, however, the most instructive approach has been deemed to be an explicit exact relativistic calculation of work done by charges on other charges in a system.

Consider the configuration shown in Fig. 1. Two charges, $q_1$ and $q_2$, are caused to revolve clockwise in a circular orbit at constant speed, such that $\beta = \omega b/c$. Since the charges are on opposite sides of the circular orbit, their velocities are always equal and opposite.

Also shown in Fig. 1 is the position of the charge $q_1$ at the retarded time $t' = t - |\mathbf{R}'|/c$, where $\mathbf{R}'$ is the displacement vector from $q_1(t')$ to $q_2(t)$. The field at $q_2(t)$ was produced by $q_1$ at the retarded time $t'$. From the position of $q_2(t)$, the angle between the position of $q_1$ at the retarded time $t'$ and at the present time $t$ is $\alpha$. From the origin $O$ of the $x$-$y$ coordinate system, at the center of the circular orbit, the angle between $q_1(t')$ and $q_1(t)$ is $\phi$.

If the charges $q_1$ and $q_2$ both equal $q$, the net dipole moment of this system is the sum of two equal but opposite dipole moments, $+qb$ and $-qb$, and is always zero. A multipole expansion shows that the charge distribution in this case is a linear combination of a monopole moment, $+2q$, and a quadrupole moment. The only nonvanishing components of the quadrupole moment tensor are the diagonal components, $Q_{xx} = 4qb^2$, $Q_{yy} = Q_{zz} = -2qb^2$. The total power radiated by this quadrupole is [11]

$$P_q = \frac{\omega^6}{360c^5}\sum_{j,k}|Q_{jk}|^2 = \frac{q^2 b^4 \omega^6}{15c^5} = \frac{\beta^2}{10}P_L, \qquad (2)$$

where $P_L$, defined in Eq. (1), is the Larmor power that would be radiated by each of the accelerated charges without the other.

Thus, if the net dipole moment vanishes ($q_1 = q_2$), the total power radiated by the simple system of charges in Fig. 1 is a factor of $\beta^2/10$ less than the dipole power that would be radiated by each charge alone. To find out why the dipole radiation vanishes for this simple system, we calculate the work done on each charge by the other. By symmetry, the work done by $q_1$ on $q_2$ must equal the work done by $q_2$ on $q_1$. Therefore, we only need to consider the field at $q_2(t)$ produced by $q_1$ at the retarded time $t'$. We only need to consider the electric field, since the magnetic field can do no work. And we only need to consider the component of the electric field in the direction of motion of $q_2$ (the $y$ direction in Fig. 1), since that is the only component that can do work at time $t$. By rotational symmetry and constant angular velocity, whatever rate of work is done on $q_2$ at time $t$ is the same for all time.

### III. CALCULATION

The electric field at $q_2(t)$ from the charge $q_1(t')$ is [12]

$$\mathbf{E}(\mathbf{R}',t) = \frac{q_1(1-\beta'^2)(\mathbf{n}'-\boldsymbol{\beta}')}{K'^3 R'^2} + \frac{q_1 \mathbf{n}'\times[(\mathbf{n}'-\boldsymbol{\beta}')\times\dot{\boldsymbol{\beta}}']}{cK'^3 R'}. \quad (3)$$

All the quantities on the right-hand side of Eq. (3) are evaluated at the retarded time $t'$. Here $\mathbf{n}'$ is the unit vector from $q_1(t')$ to $q_2(t)$, $K'$ is defined as $1-\mathbf{n}'\cdot\boldsymbol{\beta}'$, and $\dot{\boldsymbol{\beta}}'$ is the time derivative of $\boldsymbol{\beta}'$, which is the velocity normalized to $c$ of the charge $q_1$ at $t'$. The first term on the right-hand side of Eq. (3) is the "velocity field," and the second term is the "acceleration field," which depends linearly on $\dot{\boldsymbol{\beta}}'$.

From Fig. 1 and Eq. (3), the $y$ component of $\mathbf{E}(\mathbf{R}',t)$ is

$$E_y(\mathbf{R}',t) = \frac{q_1(1-\beta^2)(-\sin\alpha + \beta\cos\phi)}{K'^3 R'^2} \\ + \frac{q_1 \dot{\beta}'[\beta + \sin(\phi-\alpha)]\cos\alpha}{K'^3 R' c}, \qquad (4)$$

where we have used $|\boldsymbol{\beta}'| = \beta$, one of the simplifications made possible by constant angular velocity.

Further simplifications of Eq. (4) are possible through other exact constant relationships. For example, at an angular velocity $\omega = \beta c/b$, the charge $q_1$ traverses the arc length $b\phi$

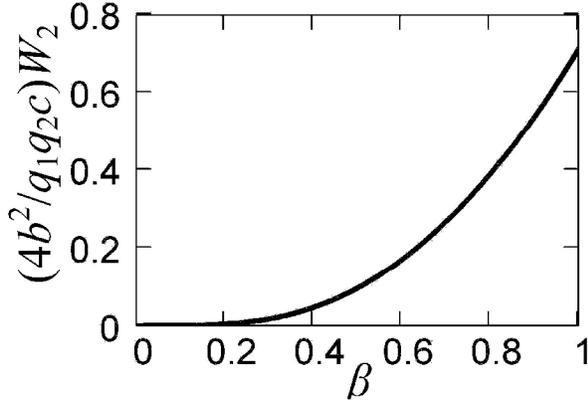

FIG. 2. Rate of work done by $q_1$ on $q_2$, in units of $q_1 q_2 c / 4b^2$, vs. $\beta$ for the configuration in Fig. 1, from Eq. (10).

during the time $R'/c$ that it takes light to travel from $q_1(t')$ to $q_2(t)$, so that

$$\beta = b\phi/R' = (2b/R')\alpha , \tag{5}$$

where the geometrical relationship $\phi = 2\alpha$ has been used. Other exact constant relationships used to simplify Eq. (4) are

$$\beta = \alpha \sec \alpha , \tag{6}$$

$$\dot{\beta}' = c\beta^2/b , \tag{7}$$

$$K' = 1 + \beta \sin \alpha . \tag{8}$$

The rate of work done by $q_1$ on $q_2$ at $t$ is

$$W_2 = q_2 \beta c E_y(R',t) . \tag{9}$$

Combining Eqs. (4) through (9), and defining a new constant $\varepsilon \equiv \sin \alpha$, gives the exact rate of work done by $q_1$ on $q_2$ at $t$ as

$$W_2 = \frac{q_1 q_2 c}{4b^2} \left[ \frac{\beta(1-\beta^2)[\beta(1-2\varepsilon^2)-\varepsilon] + 2\beta^3(\beta+\varepsilon)(1-\varepsilon^2)}{(1-\varepsilon^2)(1+\beta\varepsilon)^3} \right]. \tag{10}$$

Although the power radiated by $q_1$ approaches infinity in the limit $\beta \to 1$, in this limit the rate of work done by $q_1$ on $q_2$ approaches the finite limit $W_2 \to 0.714 q_1 q_2 c / 4b^2$, as shown in Fig. 2.

Equation (10), plotted in Fig. 2, is exact for all orders of $\beta$. More illuminating, however, is an expansion of Eq. (10) in orders of $\beta$. From Eq. (6), $\varepsilon$ may be expanded as

$$\varepsilon \equiv \sin \alpha = \beta - \frac{2}{3}\beta^3 + \frac{4}{5}\beta^5 + O(\beta^7) . \tag{11}$$

Substituting this expansion of $\varepsilon$ into Eq. (10) gives the rate of work done by $q_1$ on $q_2$ to order $\beta^6$ as

$$W_2 = \frac{2}{3} \frac{q_1 q_2 c}{b^2} \left[ \beta^4 - \frac{14}{5}\beta^6 + O(\beta^8) \right]. \tag{12}$$

Since the acceleration of the charges is $a = \beta^2 c^2 / b$, the first term in the expansion of Eq. (12) is

$$W_2 \approx 2 q_1 q_2 a^2 / 3c^3 . \tag{13}$$

## IV. RESULTS

If $q_1$ and $q_2$ both equal $q$, the net dipole moment vanishes. Then, to order $\beta^4$, Eqs. (12) and (13) show that the rate of work done by each charge on the other is equal to the nonrelativistic Larmor power $P_L$ that would be radiated by each charge in the absence of the other. (As found by Liénard in 1898 [13], the relativistic radiated power is greater than $P_L$ by a factor $(1-\beta^2)^{-2}$.) To order $\beta^4$ in Eq. (10), the rate of work done by the "acceleration field" of $q_1$ on $q_2$ is $3P_L/2$, and by the "velocity field" is $-P_L/2$, in this simple system having zero net dipole moment.

This system of two charges, having zero dipole moment, produces no dipole radiation. Each charge does work on the other at the same rate (the Larmor power) that work is done on it. Suppose the system is a nonrelativistic, frictionless spinning wheel, with two equal point charges fixed on opposite sides of the circumference of the wheel. The wheel can only lose kinetic energy by producing radiation. The only radiation it can produce, however, is from the quadrupole and higher multipole moments. Therefore from Eq. (2), the mechanical power needed to sustain a constant angular velocity $\omega$ of the frictionless wheel with zero dipole moment is less than $P_L$ by a factor 10 times smaller than $\beta^2 = (\omega b/c)^2 \ll 1$.

If the charges $q_1$ and $q_2$ are equal but opposite, then the dipole moment of this system is twice the dipole moment of the positive charge alone. Then the field of $q_1$ opposes the motion of $q_2$, and does negative work at a rate $W_2 \approx -P_L$. The total power radiated from this system is $4P_L$. Therefore, the mechanical power needed to sustain the constant angular velocity of this frictionless system must be $4P_L$ or $2P_L$ per charge, since that much power is carried away from the system by radiation.

The results concerning dipole radiation from the system of charges $q_1$ and $q_2$ shown in Fig. 1, are summarized as follows. The magnitude of the dipole moment is $|q_2 - q_1|b$. The total dipole power radiated from the system is proportional to the square of the dipole moment [14],

$$P_r \approx 2(q_2 - q_1)^2 a^2 / 3c^3 . \tag{14}$$

The mechanical power needed to sustain a constant angular velocity of the system is $P_m \approx P_r$.

## V. DISCUSSION

The results of this paper apply to charge (or antenna) source dimensions much smaller than the characteristic radiation wavelength. In Fig. 1, the source dimension is $2b$, and the characteristic wavelength is $2\pi c/\omega = 2\pi b/\beta \gg 2b$. Although the expression for the electric field in Eq. (3) is exact everywhere, including the source region, the field within the source region cannot even be considered to be in the "near (static) zone," which begins only at a distance much greater than $2b$ [15]. In this discussion, therefore, "<u>emitted</u> dipole power" refers to the rate of loss of electromagnetic energy from a charge, whether by radiation or by work done on other charges, and "<u>absorbed</u> dipole power" refers only to the rate of work done by the fields of charges in the source region, and not to any radiation process.

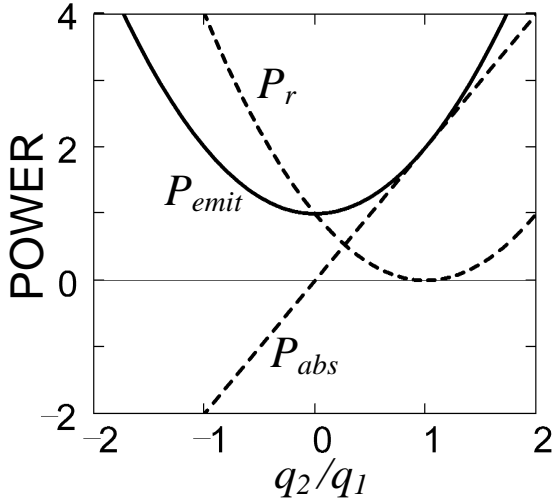

FIG. 3. Total Larmor power emitted, radiated, and absorbed, relative to emitted Larmor power of $q_1$, vs. $q_2/q_1$ for the configuration in Fig. 1, from Eqs. (14) – (17).

Not all the electromagnetic power emitted by the charges in a system necessarily radiates away from the system. Some of the power emitted by the charges may get absorbed by other charges. The total dipole power radiated from the system is the power emitted by the charges less the power absorbed,

$$P_r = P_{emit} - P_{abs}, \qquad (15)$$

where the power absorbed is given by Eq. (13) and symmetry as

$$P_{abs} \approx W_1 + W_2 = 2W_2 \approx 4q_1 q_2 a^2 / 3c^3. \qquad (16)$$

Combining Eqs. (14) to (16) shows that the power emitted by the system of Fig. 1,

$$P_{emit} = P_r + P_{abs} \approx 2(q_1^2 + q_2^2)a^2 / 3c^3, \qquad (17)$$

is just the sum of the Larmor power emitted by each charge. Figure 3 shows the total power emitted, radiated, and absorbed by the system of Fig. 1 as a function of the ratio of charges, $q_2/q_1$. The emitted power (solid curve in Fig. 3) is the sum of the radiated and absorbed powers (dashed curves), as given by Eq. (17).

The dipole power emitted by $q_1$ alone is $P_{1e} \approx 2q_1^2 a^2 / 3c^3$, but the power radiated by $q_1$ is $P_{1r} = P_{1e} - W_2 \approx 2q_1(q_1 - q_2)a^2 / 3c^3$. The remainder of the power emitted by $q_1$ is spent doing work on $q_2$ at a rate $W_2$. Similarly, the dipole power emitted by $q_2$ alone is $P_{2e} \approx 2q_2^2 a^2 / 3c^3$, but the power radiated by $q_2$ is $P_{2r} = P_{2e} - W_1 \approx 2q_2(q_2 - q_1)a^2 / 3c^3$. The remainder of the power emitted by $q_2$ is spent doing work on $q_1$ at a rate $W_1 = W_2$. The total power radiated from the system is just the sum of the power radiated from each charge, $P_r = P_{1r} + P_{2r}$, as given by Eq. (14).

So, what happens to the dipole electromagnetic power emitted by each accelerating charge in a system with zero net dipole moment? The analysis gives a clear answer. Each charge in such a system emits power at the Larmor rate, as though it were alone. But all of the emitted power gets absorbed by doing work on other charges. None of the emitted dipole power gets radiated from a system with zero net dipole moment.

This conclusion is very different from supposing that electromagnetic power does not even get emitted from accelerating charges in a system with zero dipole moment, or that the dipole power somehow gets switched off at the source. That supposition might suggest, for example, that there is no broadening of spectral lines from radiative reaction in a system with zero net dipole moment.

Neither is it helpful to suppose that the dipole power emitted from charges in a system with zero dipole moment is radiated, but that the radiation is cancelled by destructive interference in the far (radiation) zone. That supposition might suggest that positive energy densities of the radiation fields of each of the charges are somehow annihilated in the far zone.

The same conclusions hold true for a system of antennas, which is just equivalent to a system of accelerating charges. The electromagnetic field in the immediate neighborhood of an antenna was calculated in [16]. For example, consider two identical parallel dipole antennas, each emitting the same dipole power $P_{ant}$ at the same frequency. If the two antennas are exactly out of phase, then the dipole power emitted by each is absorbed by the other, and no net dipole power is radiated from the system. The power of each antenna is absorbed by the electric field of each antenna doing work on the current in the other at the rate $P_{ant}$. In that case, the dipole radiation resistance of each antenna is zero [17].

Next, consider the same two dipole antennas, but each emitting $P_{ant}$ exactly in phase. Then the electric field of each antenna does work against the current in the other at the rate $P_{ant}$. In that case, the dipole radiation resistance is doubled in each antenna, and the dipole power radiated from the pair of antennas is four times the power that would be radiated from each antenna in the absence of the other.

In the introduction to this paper, the approximate Larmor power formula was derived from first principles. That derivation raises some interesting fundamental questions about the mass-energy of electromagnetic fields and about the radiative reaction force.

The energy density of a static electric field $E$ is $E^2/8\pi$ [18]. This energy density has an equivalent mass density of $E^2/8\pi c^2$. For example, the equivalent mass of the electric field of a static charge $q$ outside a radius $R$ is $q^2/2Rc^2$. This mass is a real physical quantity. The distributed mass of an electromagnetic field even attracts other masses gravitationally [19]. The mass of the electromagnetic field differs from the mass of the charged particle that produced it, however, in that it is distributed over space, and not localized at the particle.

It is the finite speed $c$ of disturbances to the distributed mass of electromagnetic fields that gives rise to radiation from accelerated charges. From the derivations in this paper, it is easy to see that in the limit $c \to \infty$, the power radiated by an accelerated charge vanishes. In this unphysical limit, the static field of a charge would move rigidly everywhere with the charge, and would not do work on its self field. And in this unphysical limit, presumably the mass of the field would effectively be included in the mass of the charge, because any acceleration of the mass of the charge would produce an instantaneous acceleration of the entire mass of the distributed field.

Since $c$ is not infinite, the mass of the electromagnetic field is easily distinguished from the mass of the charge that produced it. That is why a greater force must be applied to a charged particle to produce the same acceleration as an identical uncharged particle. Both the mass of the charged particle and the mass of the electromagnetic field beyond the charged particle must be accelerated.

Another way of looking at the fact that a greater force is needed to produce the same acceleration of a charged particle as an identical uncharged particle is to recognize that the charged accelerated particle produces radiation, and the uncharged particle does not. Conservation of energy therefore requires that more work be done on the charged particle than on the uncharged particle to produce the same acceleration. The additional force that is needed to produce the same acceleration of a charged particle as an identical uncharged particle is known as the radiative reaction force [20]. In the model presented in this paper, the radiative reaction force is provided to the system by the mechanical force $P_m/\beta c$ that keeps the system rotating at constant velocity, even as energy is radiated away.

Since the dipole power radiated by a source depends on the work done by charges on other charges in the source region, one might expect that modifications to the source region might affect the radiated power just as modifications to the source would. For example, an experiment was recently proposed to demonstrate that "almost all far-field power can be provided by evanescent waves" by modifying the index of refraction in the neighborhood of a radiating dipole [21].

As another example, certain laser-target plasmas in vacuum have no electric dipole moment. But because just a part of the poloidal electron current passes through an overdense plasma plume, where the electromagnetic power emitted by the current is converted to electroacoustic waves, dipole power is radiated from the other part of the current, even though the poloidal current has no electric dipole moment [22].

An analogy to this process would be if one of the two charges in Fig. 1 were shielded. Then, even if the charges were equal, so that the system had no dipole moment, the unshielded charge would produce dipole radiation because it could do no work on the shielded charge.

In conclusion, an exact calculation of high-order relativistic corrections of the electric fields of dipoles in their source regions accounts for dipole power radiated in the far field. The Larmor power emitted by each dipole in a system of zero dipole moment is absorbed by doing work on the other dipoles in the system. And in a system of *N* parallel identical dipoles oscillating in phase, each of the dipoles must work against the retarded fields of all the other dipoles in the system. Each dipole therefore radiates *N* times as much power as a solitary dipole would, and the system radiates $N^2$ times as much as a solitary dipole.

**ACKNOWLEDGMENT**

The author is grateful to Neal Carron for helpful discussions on dipole antennas.